\documentclass[conference, compsoc, 10pt]{IEEEtran}

\usepackage{amsmath}

\usepackage{amssymb,amsfonts}
\usepackage{algorithmic}
\usepackage{graphicx}
\usepackage{textcomp}
\usepackage{standalone}
\usepackage{xcolor}
\def\BibTeX{{\rm B\kern-.05em{\sc i\kern-.025em b}\kern-.08emT\kern-.1667em\lower.7ex\hbox{E}\kern-.125emX}}
\usepackage[utf8]{inputenc}
\usepackage{booktabs} 
\usepackage{graphicx}
\usepackage{subfigure}
\usepackage{units}
\usepackage{blindtext}
\usepackage{multirow}
\usepackage{xcolor,xspace}
\usepackage{amsmath}
\usepackage{paralist}
\usepackage{mdframed}
\usepackage{pict2e}
\usepackage{listings}
\usepackage{color}
\usepackage{threeparttable}
\usepackage{array}
\usepackage{booktabs}
\usepackage{pifont}
\usepackage{url}
\usepackage{comment}
\usepackage{amsmath}
\usepackage{lipsum}
\usepackage{arydshln}
\usepackage{flushend}
\usepackage{latexsym}
\usepackage{enumitem}
\usepackage{fancyhdr}
\usepackage[english]{babel}
\usepackage{tabularx}
\input{glyphtounicode}

\usepackage[ruled]{algorithm2e}
\usepackage{algorithmic}

\usepackage{url}
\usepackage[
    n,
    advantage,
    operators,
    sets,
    adversary,
    landau,
    probability,
    notions,
    logic,
    ff,
    mm,
    primitives,
    events,
    complexity,
    asymptotics,
    keys]{cryptocode}

\usepackage{tikz}                    
\usetikzlibrary{arrows, patterns, automata, positioning}

\newcommand{\ignore}[1]{}

\newcommand{\prv}{{\ensuremath{\sf{\mathcal Prv}}}\xspace}
\newcommand{\vrf}{{\ensuremath{\sf{\mathcal Vrf}}}\xspace}

\newcommand{\chal}{{\textit{Chal}}\xspace}
\newcommand{\RA}{{\textit{RA}}\xspace}
\newcommand{\PoX}{{\textit{PoX}}\xspace}
\newcommand{\CFA}{{\textit{CFA}}\xspace}

\newcommand{\acron}{\textit{{Spec\CFA}}\xspace}
\newcommand{\data}{\textit{{BlockMem}}\xspace}

\renewcommand\adv{\ensuremath{\sf{\mathcal Adv}}\xspace}

\newcommand{\cflog}{\ensuremath{CF_{Log}}\xspace}
\newcommand{\ptr}{\ensuremath{\mathcal{C}F_{Size}}\xspace}

\newlist{myenumerate}{enumerate}{1}
\setlist[myenumerate]{
  label=\arabic*),
  align=left,
  leftmargin=*,
  nosep,
}

\AtBeginDocument{%
  \providecommand\BibTeX{{%
    Bib\TeX}}}

\begin{document}

\title{\acron: Enhancing Control Flow Attestation/Auditing\\ via Application-Aware Sub-Path Speculation}

\author{
\IEEEauthorblockN{Adam Caulfield, Liam Tyler, and Ivan De Oliveira Nunes}
\IEEEauthorblockA{Rochester Institute of Technology, USA}
}

\maketitle

\begin{abstract}
At the edge of modern cyber-physical systems, Micro-Controller Units (MCUs) are responsible for safety-critical sensing/actuation.
However, MCU cost constraints rule out the usual security mechanisms of general-purpose computers.
Thus, various low-cost security architectures have been proposed to remotely verify MCU software integrity.
Control Flow Attestation (\CFA) enables a Verifier (\vrf) to remotely assess the run-time behavior of a prover MCU (\prv), generating an authenticated trace of all of \prv control flow transfers (\cflog). Further, Control Flow Auditing architectures augment \CFA by guaranteeing the delivery of evidence to \vrf.

Unfortunately, a limitation of existing \CFA lies in the cost to store and transmit \cflog, as even simple MCU software may generate large traces.
Given these issues, prior work has proposed static (context-insensitive) optimizations. However, they do not support configurable program-specific optimizations.
In this work, we note that programs may produce unique predictable control flow sub-paths and argue that program-specific predictability can be leveraged to dynamically optimize \CFA while retaining all security guarantees.
Therefore, we propose \acron: an approach for dynamic sub-path speculation in \CFA. 
\acron allows \vrf to securely speculate on likely control flow sub-paths for each attested program. At run-time, when a sub-path in \cflog matches a pre-defined speculation, the entire sub-path is replaced by a reserved symbol. \acron can speculate on multiple variable-length control flow sub-paths simultaneously. We implement \acron atop two open-source control flow auditing architectures: one based on a custom hardware design~\cite{acfa} and one based on a commodity Trusted Execution Environment (ARM TrustZone-M)~\cite{iscflat}. In both cases, \acron significantly lowers storage/performance costs that are critical to resource-constrained MCUs.
\end{abstract}

\section{Introduction}
\label{sec:intro}

\begin{figure}[!b]
\begin{minipage}{\columnwidth}
  \vspace{-1em}
  \rule{\linewidth}{0.4pt}
  \noindent\small{\textbf{To appear: \textit{The 40th Annual Computer Security Applications Conference (ACSAC'24)}}}.
\end{minipage}
\end{figure}

Micro-Controller Units (MCUs) are part of cyber-physical systems and implement the \textit{de-facto} interface between the physical and digital worlds. Therefore, they are relied upon to implement safety-critical sensing and actuation tasks~\cite{opensyringe, noman2018design, simple}.
However, due to energy, size, and cost constraints, MCUs lack security features common to general-purpose computers. In particular, they usually run software at bare-metal, lacking Memory Management Units (MMUs), fine-grained inter-process isolation, or strong privilege separation.
Given their importance to modern systems and lack of security features, MCUs have become attractive targets for attacks~\cite{nafees2023smart, kayan2022cybersecurity}.

In this context, Remote Attestation (\RA)~\cite{simple,ra_mag,smart,vrased,tytan,trustlite,hydra,rata,Sancus17,pistis,reserve,sacha,ammar2018wise,surminski2023dma,surminski2021realswatt,scraps,KeJa03,SPD+04,SLS+05,SLP08,PFM+04,KKW+12,SWP08,li2011viper,surminski2023scatt} 
and Proofs of Execution (\PoX)~\cite{apex,asap} were proposed as inexpensive means to detect software compromises on MCUs. In a typical \RA protocol, a \textit{Verifier} (\vrf) aims to determine if the software state of a remote \textit{Prover} MCU (\prv) is trustworthy.
\RA produces an authenticated and unforgeable ``snapshot'' of \prv's current software image. \PoX builds atop \RA to also prove that a particular function within this software image has been executed in a timely manner.

On the other hand, neither \RA nor \PoX offers evidence about the order in which instructions have been executed.
Thus, out-of-order execution attacks, including control flow hijacking~\cite{schuster2015counterfeit, evans2015control}, Return-Oriented Programming (ROP)~\cite{shacham2007geometry}, and Jump-Oriented Programming (JOP)~\cite{bletsch2011jump}, remain oblivious to \vrf  with \RA and \PoX.

While Control Flow Integrity (CFI)~\cite{Abadi2009,FC08,hafix,DSL14,ge2017griffin,stackguard,mishra2023procrastinating} can
detect some attacks locally at \prv, they do not provide \vrf with evidence about the malicious control flow path taken, precluding analysis of the anomalous behavior~\cite{sok_cfa_cfi}. Therefore, Control Flow Attestation (\CFA)~\cite{iscflat,lofat,tinycfa,oat, dessouky2018litehax,zeitouni2017atrium,scarr, recfa,cflat,wang2023ari,dessouky2019chase,abera2019diat} was proposed to provide precise evidence of the execution's control flow path to a remote \vrf. Until recently, \CFA was limited by the fact that a fully compromised \prv could ignore \vrf requests or refuse to deliver a \cflog that indicates a compromised state. This problem is recognized and addressed in recent work~\cite{acfa} by incorporating reliable delivery of \cflog as part of the \CFA's root of trust (RoT) functionality to enable ``control flow auditing'' as opposed to best-effort \CFA. We revisit \CFA details in Sec.~\ref{sec:background}.

Regardless of specifics, {\it CF}-Attestation/-Auditing~\footnote{In the remainder of this paper (unless explicitly stated) we use \CFA to refer to both Control Flow Attestation and Control Flow Auditing schemes.}
requires the storage of \cflog and its eventual transmission to \vrf, which becomes a bottleneck on the resource-constrained \prv.
For this reason, earlier \CFA methods~\cite{lofat,zeitouni2017atrium,cflat} proposed to build \cflog as a hash-chain to compress the sequence of control flow transfers into a single hash digest.
While this approach minimizes storage/transmission costs, it requires \vrf to derive the exact control flow path that could have led to the received hash digest (for the entire execution). The complexity of this task grows exponentially with the number of control flow transfers, leading to the well-known path explosion problem~\cite{aliasing, baldoni2018survey}.
Due to path explosion, more recent \CFA approaches opt to store \cflog verbatim~\cite{iscflat,tinycfa,scarr,recfa} or in compressed form without loss of information~\cite{oat,dessouky2018litehax}. Consequently, the aforementioned costs limit their applicability to small operations~\cite{oat,tinycfa}.
While simple optimizations to reduce \cflog (such as replacing simple loops with counters~\cite{acfa,tinycfa,cflat}; or storing decision bits instead of destination addresses for simple conditional branches~\cite{oat,dessouky2018litehax}) have been proposed, they are static. In other words, they are hard-coded on the \CFA RoT and unaware of the application being attested in a ``one size fits all'' fashion.

Contrary to static approaches, our work is motivated by two key premises:
{\bf (1)} Expected control flow sub-paths are often predictable and repetitive. Therefore, with knowledge of the attested program's expected behavior (e.g., based on prior execution observation), one can learn likely sub-paths with reasonable accuracy;
{\bf (2)} Anomalous and malicious \cflog-s are rare. While \vrf should still receive {\bf \underline{all}} control flow information, attacks do not happen as often as benign executions. Therefore, it should be possible to replace expected benign sub-paths with small reserved symbols to indicate that an entire benign/expected sub-path has occurred (instead of filling \cflog with redundant data).
Based on these premises, we propose \acron: an approach to enable dynamic sub-path speculation policies in \CFA. 

\acron provides \vrf with the ability to speculate on expected control flow transfers for the attested operation. If speculations match an execution sub-path taken by \prv at run-time, the entire matching sub-path is replaced by a reserved symbol, greatly reducing \cflog size without loss of information.
\acron allows simultaneous speculation on multiple variable-sized sub-paths for a given attested operation.

Unsurprisingly, the ability to dynamically speculate on sub-paths also opens attack vectors that (if left unprotected) could be exploited by an adversary (\adv) to forge/spoof execution traces. This leads to non-trivial challenges that must be overcome to realize \acron securely. Therefore, we also specify and implement architectural measures to guarantee that \acron retains the same security as the underlying \CFA architectures while achieving performance gains.

We note that existing \CFA is either based on Trusted Execution Environments (TEEs) or custom hardware support.
To demonstrate \acron's generality, we implement it atop one representative of each category, namely: {\bf (1)} ACFA~\cite{acfa} which targets lowest-end MCUs and employs custom hardware support; and {\bf (2)} ISC-FLAT~\cite{iscflat} which targets ``off-the-shelf" MCUs, leveraging a commodity TEE (TrustZone for ARM Cortex-M). In the former, \acron is instantiated as a custom hardware design. In the latter, \acron is implemented within the TEE's trusted world. We choose these architectures due to their public availability. Nonetheless, we believe \acron's concepts to be broadly applicable to any \CFA architecture.

Finally, we also propose and evaluate several approaches to determine effective path speculations.
We consider and compare automated approaches based on previously observed \cflog-s and static analysis, in addition to manual analysis of expected control flow paths. In several cases, \acron leads to order-of-magnitude improvements in storage, bandwidth, and communication latency while retaining all \CFA guarantees. We make \acron prototype implementation publicly available at~\cite{specrepo}.
%
%
\section{Background \& Related Work}
\label{sec:background}

\subsection{Remote Attestation (\RA)}\label{sec:bg_ra}

\RA is a challenge-response protocol in which a \vrf aims to check the software image currently installed on \prv, i.e., the content of \prv's program memory (PMEM). 
A typical \RA protocol is performed as follows:

\begin{myenumerate}
  \item \vrf sends \prv a unique cryptographic challenge (\chal).
  \item After authenticating \vrf's request containing \chal, an RoT in \prv computes an authenticated integrity-ensuring function over PMEM and \chal to produce a response ($H$).
  \item \prv sends $H$ to \vrf.
  \item \vrf compares $H$ to its expected value.
\end{myenumerate}


Step (2) can be implemented using a Message Authentication Code (MAC) or digital signature.
The secret key used in this operation must be securely stored by the RoT on \prv to ensure it is inaccessible to untrusted software.

\RA architectures are usually classified into three categories: software-based, hardware-based, or hybrid, depending on how their RoT is implemented.
Software-based \RA~\cite{simple,pistis,ammar2018wise,surminski2021realswatt,KeJa03,SPD+04,SLS+05,SLP08} does not rely on specialized hardware but relies on strong assumptions about \adv and the system.
Hardware-based approaches~\cite{Sancus17,sacha,PFM+04,KKW+12, SWP08} rely on support from dedicated hardware, TPMs~\cite{tpm}, or instruction set features~\cite{sgx} to achieve attestation with stronger security guarantees at a higher cost. 
Hybrid \RA schemes~\cite{smart,vrased, tytan, trustlite,li2011viper} combine hardware and software to achieve security guarantees comparable to hardware-based \RA while minimizing the hardware cost.
Hybrid approaches perform the \RA measurement in software while protecting its execution and cryptographic key(s) through custom hardware.

\subsection{Control Flow Attestation/Auditing (\CFA)}
\label{subsec:background_cfa}


\CFA~\cite{sok_cfa_cfi,lofat,tinycfa,oat,dessouky2018litehax,zeitouni2017atrium,scarr,recfa,cflat,wang2023ari,dessouky2019chase,geden2019hardware} augments \RA to detect control flow attacks. 
In addition to proving if the correct software image is installed on \prv, \CFA produces an authenticated log (\cflog) of all executed control flow transfers. Existing \CFA methods use either (1) binary instrumentation and TEE support~\cite{oat, scarr, recfa, cflat}; or (2) custom hardware modifications~\cite{acfa,lofat,tinycfa,dessouky2018litehax,zeitouni2017atrium,dessouky2019chase} to securely create and store \cflog. After execution, \prv's RoT MACs/signs \cflog along with the \RA evidence to produce $H$. Upon receiving \cflog and $H$, \vrf inspects their contents to detect control flow attacks. 

Early \CFA techniques used hash-chains to return a single hash~\cite{lofat,cflat} (or multiple hashes~\cite{zeitouni2017atrium}) of \cflog to \vrf.
To verify the execution, \vrf checks if the received hash digest corresponds to a valid path. 
Although this approach reduces \cflog to a small fixed size, it is limited in its scalability to more complex software, 
as \vrf might face path explosion when attempting to produce a complete set of valid paths~\cite{aliasing, baldoni2018survey}.
Because of this, more recent \CFA designs produce {\it verbatim} \cflog-s that include the destination address of each control flow transfer.
{\it Verbatim} \cflog-s ease verification, but storage and transmission of \cflog become challenging in branch-intensive attested programs.
While one could consider standard compression algorithms (e.g., based on Huffman codes~\cite{huffman}) to reduce \cflog, these algorithms are too heavy to run locally on a resource-constrained \prv. Instead, prior work has introduced simpler strategies to reduce \cflog.
LiteHAX~\cite{dessouky2018litehax} is a hardware-based approach that records a reduced-sized bitstream of \cflog. LiteHAX logs a single bit for each direct jump/call and conditional branch (1/0 if the branch was/was not taken, respectively). As indirect branches can have multiple destinations, their full address is recorded in the bitstream. 
OAT~\cite{oat} records branch destinations in a similar manner to LiteHAX and reduces the bitstream size further by producing a hash-chain of return addresses instead of adding them to the bitstream. 
ARI~\cite{wang2023ari} follows the same logging scheme as OAT while only recording the control flow transfers that enter, exit, or occur within user-defined mission-critical components.

An alternative approach to encoding the \cflog as a bitstream is to continuously transmit a series of reduced-sized \cflog-s (slices) to \vrf. 
ScaRR~\cite{scarr} provides \CFA with \cflog slicing for complex systems such as cloud-based virtual machines.
ACFA~\cite{acfa} is a hardware-software co-design for low-end MCUs that uses custom hardware to actively interrupt execution to transmit fixed-sized \cflog slices.
ACFA is the first technique to enable control flow {\it auditing} by incorporating reliable evidence delivery as part of its RoT.

To our knowledge, no prior works consider the specificity of each software being attested to reduce \cflog size. For example, while utilizing loop counters is generally applicable, it offers little optimization to programs with few or complex loops. Similarly, this generality may miss out on unique program behaviors that offer better reductions. Our work bridges this gap by enabling 
\textit{secure and configurable program-specific speculations}
that lead to significant performance improvements in \CFA.


\section{\acron at a High-Level}\label{sec:overview}

As part of a \CFA request, \acron allows \vrf to specify which control flow sub-paths are expected to occur the most during the execution of the attested program.
Upon receiving an authenticated request from \vrf (along with the specified sub-paths), \acron RoT saves the \vrf-defined paths to protected memory and starts the attested execution of the requested program. As the program executes, the underlying \CFA architecture saves control flow transfers to \cflog. Whenever a matching control flow sub-path is found in \cflog, it is replaced by a short reserved symbol, indicating one occurrence of a \vrf-specified sub-path in \cflog.

\textit{\textbf{Remark 1:} as \vrf is aware of the path/symbol correspondence, \acron does not result in any loss of information in \cflog. \vrf can simply replace the symbols in the received \cflog to derive the full \cflog for verification.}

\begin{figure}[t]
  \centering
  \includegraphics[width=\columnwidth]{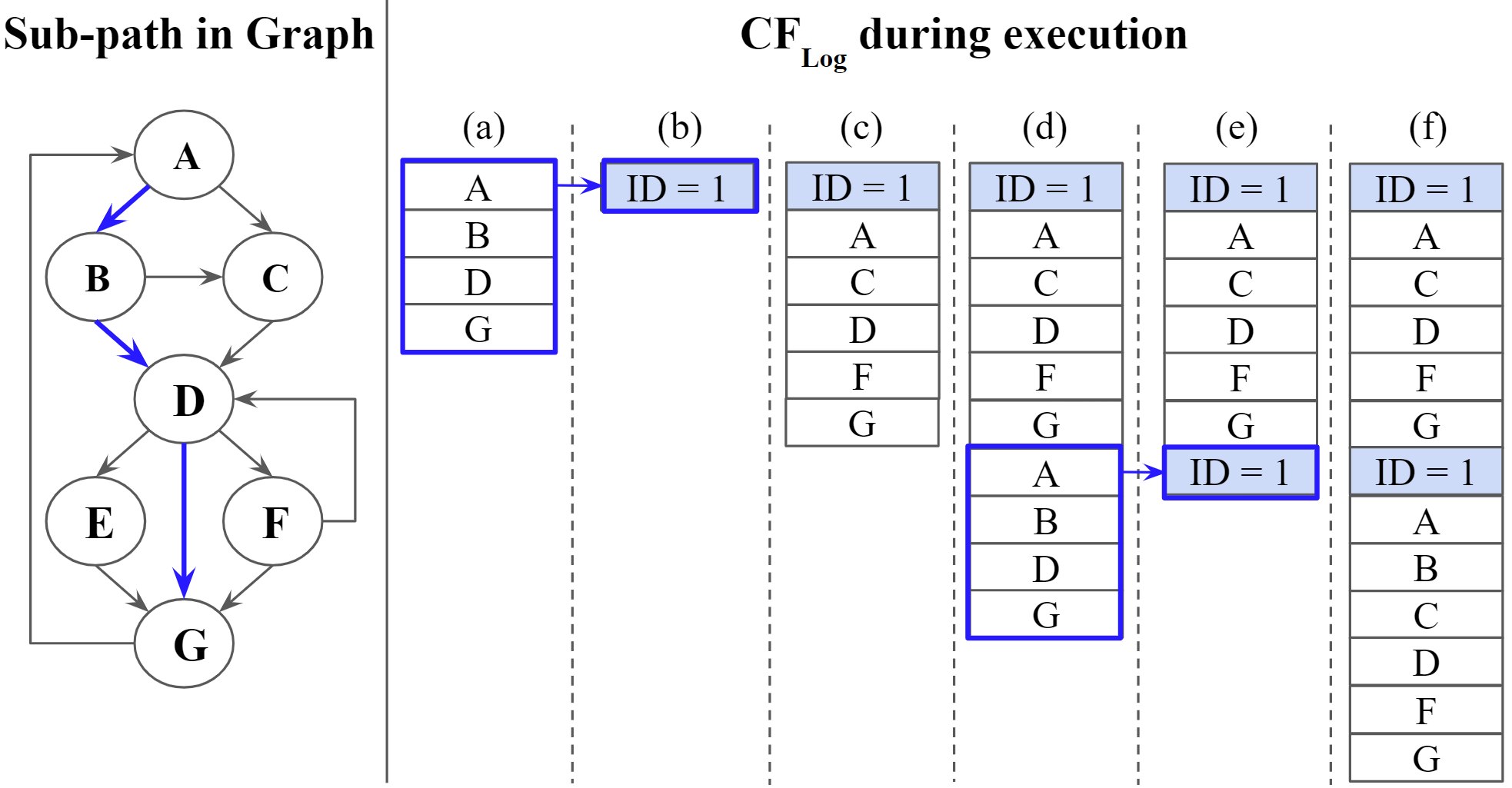}
  \caption{Example optimization made by \acron}
  \label{fig:speculate-example}
  \vspace{-1em}
\end{figure}

Fig.~\ref{fig:speculate-example} presents a high-level example of \acron's intended behavior.
In this example, \cflog is being monitored by \acron to optimize the \vrf-specified sub-path $\{A, B, D, G\}$. As the program executes, the underlying \CFA architecture appends data to \cflog after each control flow transfer. When a matching instance of the sub-path appears in \cflog (stage \texttt{(a)}), \acron detects and replaces the sub-path with its associated symbol: an ID equal to 1 in this example (stage \texttt{(b)}). As execution continues, \cflog continues to grow with new transfers (stage \texttt{(c)}) and \acron does not modify \cflog until the next match occurs (stage transition \texttt{(d)} $\rightarrow$ \texttt{(e)}). After the sub-path replacement, execution continues and new transfers are appended normally (stage \texttt{(f)}). This process continues until the end of execution.
For simplicity, this example shows an optimization of a single sub-path containing only four transfers. However, \acron supports speculation of multiple sub-paths of arbitrary length. 
As noted earlier, path speculations (e.g., $\{A, B, D, G\}$ in this example) are defined by \vrf and sent to \prv along with \CFA requests (recall Sec.~\ref{sec:background}). Importantly, \vrf is not always required to send speculation paths, but only when it decides to update a speculation strategy.

As we demonstrate in Sec.~\ref{sec:performance}, this intuitive idea results in significant performance improvements in terms of storage, bandwidth, and latency. However, from a security standpoint, \acron's design must overcome several non-trivial challenges. In particular, it must ensure that the ability to speculate on sub-paths does not lead to exploitable attack vectors (on a \prv whose software is, by assumption, potentially compromised). For instance, \adv could attempt to compromise \cflog integrity by replacing illegal paths with expected sub-path symbols. The latter can be accomplished by a variety of means, depending on \adv's strategy. Therefore, Sec.~\ref{sec:hw_based} and Sec.~\ref{sec:tz-design} detail \acron design in order to prevent any such attempt, while facilitating \acron performance gains. Then, in Sec.~\ref{sec:security}, we argue the security of the overall constructions for the two classes of \CFA considered in this work (i.e., based on custom hardware and TEEs).

\textit{\textbf{Remark 2:} 
While \acron savings do not extend to malicious or anomalous sub-paths (as \vrf would not speculate on unknown anomalies/attacks), such instances are rare. In these cases, \acron still ensures accurate detection. Conversely, during \prv's predictable and intended executions, \acron enhances performance.}

\textit{\textbf{Remark 3:} \acron inherits its support for interrupts from the underlying \CFA architecture. Some \CFA schemes~\cite{iscflat} allow but do not record external interrupts to \cflog. As such, interrupts do not affect \acron sub-path speculations. For architectures that record all interrupts in \cflog~\cite{acfa}, \acron can speculate on interrupt routine paths if included in the sub-path definition.}

\vspace{-0.25em}

\subsection{System Model \& Scope}\label{sec:adv_model}

As noted in Sec.~\ref{sec:intro}, this work focuses on resource-constrained edge embedded devices, implemented using MCUs.
These MCUs are single-core and run software atop ``bare-metal'' (executing instructions physically from program memory). They lack MMUs and strong privilege separation to support virtual memory or secure micro-kernels. This scope is aligned with the related work discussed in Sec.~\ref{sec:background}. As mentioned earlier, our work targets both types of \CFA architectures considered in the literature: those employing custom hardware designs and those based on TEEs.

Custom hardware approaches assume no reliance on pre-existent hardware features. On the other hand, they employ small hardware modifications to implement \CFA. In this case, \acron features are also implemented in hardware, as part of the \CFA RoT. This model is equivalent to prior related work, without additional assumptions. We present this version of \acron in Sec.~\ref{sec:hw_based}.

For the TEE-based version of \acron, discussed in Sec.~\ref{sec:tz-design}, our implementation instantiates unmodified ARMv8 MCUs equipped with TrustZone-M. Attested programs execute in the ``Non-Secure" World, while the ``Secure World" is trusted and used to implement the \CFA RoT (including \acron new features). This \adv model is equivalent to prior work on TEE-based \CFA, without additional assumptions. 

\subsection{Adversary (\adv) Model}\label{sec:adv_model}
\acron with custom hardware considers a strong \adv that can exploit software vulnerabilities in \prv to (1) modify any writable memory that is not explicitly protected by hardware-enforced access controls; (2) cause malicious control flow transfers; and (3) attempt to hide their malicious actions (in the form of injected code or hijacked control-flows).
Unless prevented, modifications to program memory can change instructions, and modifications to data memory can corrupt intermediate computation results and affect the program's control flow.
The TEE-based version of \acron considers that \adv can fully compromise the Non-Secure World on \prv. \adv can exploit vulnerabilities to launch code injection attacks, hijack control flow, or perform code reuse attacks. In addition, \adv can manipulate any Non-Secure World configuration registers. In each case, our \adv model remains consistent with existing \CFA architectures.
In both cases, hardware attacks that require physical access/modification to circumvent \prv hardware protections (or hardware-protected software) are out-of-scope. Protection against physical hardware attacks involves orthogonal physical access control measures~\cite{obermaier2018past}.

\section{\acron with Custom Hardware}\label{sec:hw_based}

\begin{table}[t]
\caption{Notation Summary}
\label{tab:notation}
\vspace{-0.75em}
\centering
\small
\resizebox{0.99\columnwidth}{!}{%
\begin{tabular}{|c|p{0.85\linewidth}|}
\hline
\textbf{Symbol} & \textbf{Definition} \\
\hline
PC & Program Counter (points to the current instruction). \\
\hline
$W_{en}$ & MCU write enable bit (set when writing to memory)\\
\hline
$D_{addr}$ & the MCU address being read or written from/to. \\
\hline
$DMA_{en}$ & DMA write enable bit (set when DMA writes to memory)\\
\hline
$DMA_{addr}$ & the DMA address being read or written from/to.\\
\hline
\data & Reserved memory for storing sub-path speculations\\
\hline
\cflog & log that stores attested control flow\\
\hline
$\ptr$ & current size of \cflog\\
\hline
$(src, dest)$ & source and destination of the current control flow transfer\\
\hline
$hw_{en}$ & set when \CFA module is appending \cflog with ($src, dest$) \\
\hline
$Block_{i}$ & sequence of transfers defining sub-path$_i$ speculation \\ 
\hline
$block_{base}$ & the base address of a block with respect to \data. \\
\hline
$block_{ptr}$ & points to ($block_{src}$, $block_{dest}$) being checked. \\
\hline
$block_{ID}$ & the $ID$ of a sub-path \\
\hline
$block_{len}$ & the length of the sub-path \\
\hline
$(block_{src}, block_{dest})$ & the source/destination addresses in a given sub-path entry\\
\hline
$detect_{active_i}$ & set by Block$_i$ Detect module when sub-path$_i$ has occurred\\
\hline
$active_{addr_i}$ & address of sub-path$_i$ in \cflog when $detect_{active_i}$ is set\\
\hline
$active_{ID}$ & the selected $block_{ID_i}$ when  $detect_{active_i}$ is set\\
\hline
$detect_{any}$ & a signal set when any sub-path has been detected \\
\hline
$spec_{en}$ & a signal set when any speculation (or repeat) is detected\\
\hline
$(spec_{addr}, spec_{value})$ & the location and value of the optimization and in \cflog\\
\hline
\end{tabular}
}
\vspace{-1.5em}
\end{table}

\begin{figure}[t]
  \centering
  \includegraphics[width=0.9\columnwidth]{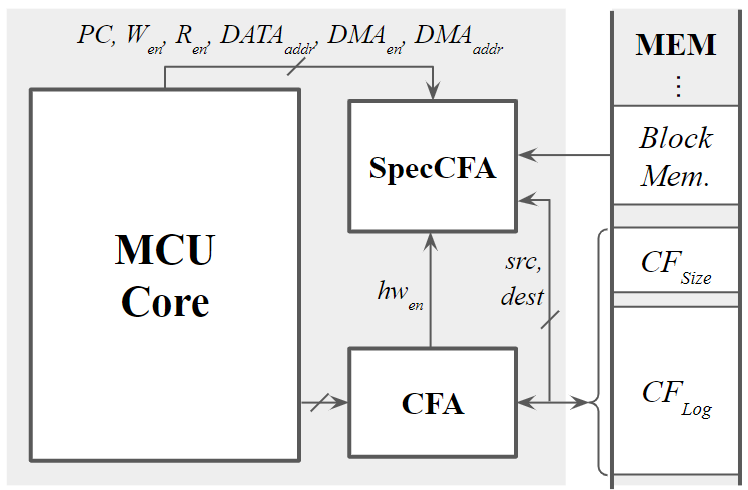}
  \caption{System overview of \acron hardware}
  \vspace{-1em}
  \label{fig:system}
\end{figure}

Fig.~\ref{fig:system} depicts a \CFA-enabled MCU architecture extended with \acron's custom hardware design. Table~\ref{tab:notation} summarizes the notation used in the rest of the paper.

The underlying \CFA architecture monitors several CPU signals at runtime. By checking the program counter (PC) as well as the $opcode$ of the currently executing instruction, it obtains control flow transfers' source and destination ($src, dest$) addresses to append them to \cflog as they occur. When writing the ($src, dest$) pair to \cflog, it sets a flag (denoted $hw_{en}$). \acron interacts with the \CFA architecture by monitoring $hw_{en}$ and ($src, dest$) being written to \cflog. In addition to these signals from the \CFA architecture, \acron hardware also monitors certain signals from the MCU to provide additional properties. $PC$ is used to determine which part of the software is executing. Signals regarding memory accesses, such as the write- and read-enable bits ($W_{en}$, $R_{en}$, respectively) indicate that the MCU is currently writing to memory ($W_{en}=1$) or reading from memory ($R_{en}=1$). Whenever a read/write occurs, the $D_{addr}$ signal contains the address of the read/written memory address. Similarly, \acron monitors Direct Memory Access (DMA) related signals ($DMA_{addr}, DMA_{en}$) to verify memory accesses performed by DMA. Whenever DMA accesses (either reads or writes) memory, $DMA_{addr}$ contains the address being accessed and $DMA_{en}=1$.

Both \acron-specific modules and the underlying \CFA architecture can modify \cflog's designated memory. In addition, both also monitor the current size of \cflog (denoted \ptr), which is incremented by the \CFA architecture whenever it appends a new entry to \cflog to track the next available memory address. 
Finally, \acron interfaces with and monitors \data, which contains memory blocks dedicated to storing the sub-path speculation definition(s), as received from \vrf in the \CFA requests.
\data, \cflog, and \ptr are stored in hardware-protected and dedicated addresses in memory that are inaccessible to untrusted software in \prv and can only be written by the \CFA and \acron modules.


\subsection{Memory Organization \& Sub-Modules}
\label{subsec:comp_details}

\begin{figure}[t]
    \centering
    \includegraphics[width=0.8\columnwidth]{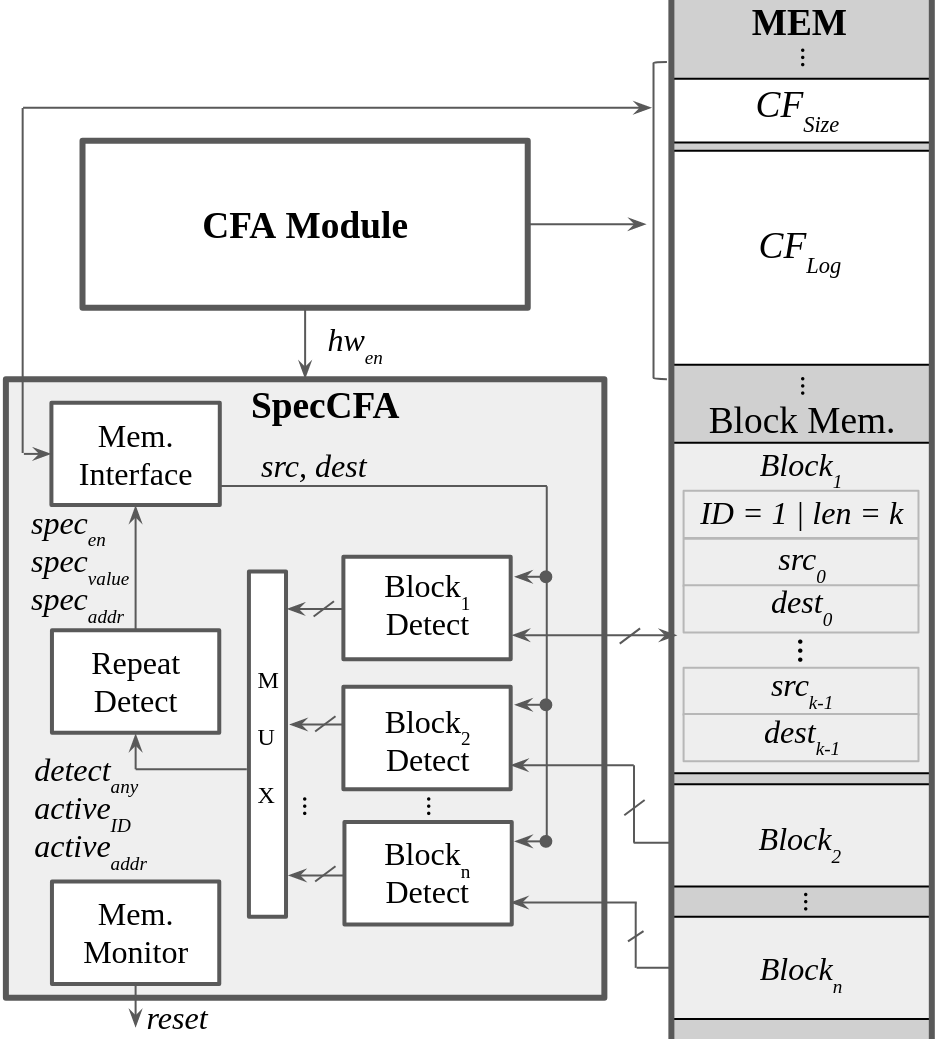}
    \caption{\acron hardware components}
    \label{fig:component-details}
    \vspace{-1.5em}
\end{figure}

In its hardware-based version, \acron is composed of four internal sub-modules, namely: Memory Interface, Block Detect module(s), Repeat Detect, and Memory Monitor.  Fig. \ref{fig:component-details} further details their interconnections.

Memory Interface performs accesses to \cflog and \ptr (in memory) to replace matching sub-paths, when detected. \acron instantiates one Block Detect module per speculated sub-path (referred to as a “block” of control flow transfers) in \data, and each instance is responsible for detecting the occurrence of their
associated sub-path in \cflog. Each Block Detect module interfaces
with \data directly to read the $ID$, $len$, and sub-path ($src$, $dest$) pairs of a given block. For the remainder of this section, we refer to the data read from \data by each Block Detect module as ($block_{ID}$,
$block_{len}$, $block_{src}$, $block_{dest}$).

Every Block Detect module compares each current ($src$, $dest$) being written to \cflog to their ($block_{src}$, $block_{dest}$) pairs in \data to determine if their sub-path has occurred. At each match, they iterate through \data to determine the next expected ($block_{src}$, $block_{dest}$) pair to be checked. When a full sub-path is detected, they output the following signals:

\begin{myenumerate}
     \item $\mathit{detect_{active}}$: a flag indicating that a sub-path has been detected.
     \item $\mathit{active_{addr}}$: the memory address of the sub-path in \cflog as an offset to the base address of \cflog.
     \item $\mathit{block_{ID}}$: the $ID$ associated with the detected sub-path.
\end{myenumerate}

All Block Detect modules’ outputs are selected by a multiplexer (MUX), using $detect_{active}$ bits as a selector. 

The Repeat Detect module monitors detected sub-paths for repeated adjacent speculations. The module replaces repeated speculations with a counter of how many times the sub-path occurred successively to further reduce \cflog size (instead of logging the same symbol multiple times). The module receives a logic $OR$ of all $detect_{active}$ signals, indicating that one of the Block Detect modules detected a sub-path. In addition, it receives $active_{ID}$ and $active_{addr}$ as outputs from the MUX, which contain the $block_{ID}$ and memory address in \cflog of the detected sub-path. Repeat Detect then compares the detected sub-path with the previous speculation and sets three output signals: $spec_{value}$, $spec_{addr}$, and $spec_{en}$. If a repeat is detected, the module increments an internal repeat counter, sets $spec_{value}$ to the value of this counter, and $spec_{addr}$ to the address of the counter in \cflog. Otherwise, $spec_{value}$ and $spec_{addr}$ are set to $active_{ID}$ and $active_{addr}$, respectively. In either case, $spec_{en}$ is set to indicate a speculation match was found.

The Memory Interface module receives $spec_{en}$ along with $spec_{value}$ and $spec_{addr}$ which contain the current match and address of the match in \cflog. When $spec_{en}$ is set, The Memory Interface then
writes $spec_{value}$ to \cflog at $spec_{addr}$ and decrements \ptr according to the size reduction due to the matching sub-path replacement.

Finally, the Memory Monitor ensures that no untrusted software on \prv can edit \data. It triggers a hardware exception whenever CPU-writes
or DMA-writes to \data are attempted unless they originate from within the \CFA RoT. To detect writes within the bounds of \data and determine whether they originate from the \CFA's RoT, Memory Monitor checks the signals:
$PC$, $W_{en}$, $D_{addr}$, $DMA_{en}$, and $DMA_{addr}$ (see Sec. \ref{subsec:hw_specs} for details).

\subsection{Sub-Path Detection Finite State Machine}
\label{subsec:states}

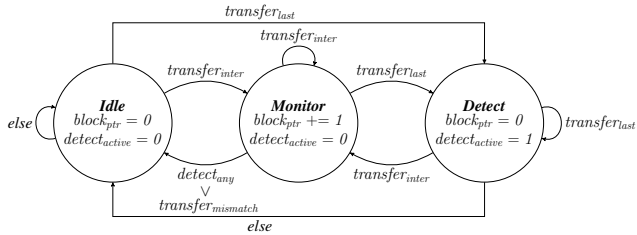
\begin{figure}[t]
\begin{center}
\vspace{1em}
\noindent\resizebox{0.99\columnwidth}{!}{%

\begin{tikzpicture}[->,>=stealth',auto,node distance=cm,semithick]
\tikzstyle{every state}=[minimum size=1.5cm]
 \tikzstyle{every node}=[font=\LARGE]

\node[state] (0,0) (I) {
        \shortstack{
        \textbf{\textit{Idle}} \\ 
        $\mathit{block_{ptr} = 0}$ \\ 
        $\mathit{detect_{active} = 0}$
        }};

\node[state] [right=2.5cm of I] (M) {
        \shortstack{\textbf{\textit{Monitor}} \\ $\mathit{block_{ptr}\mathrel{+}= 1}$ \\ $\mathit{detect_{active} = 0}$}};

\node[state] [right=2.5cm of M] (D){\shortstack{
        \textbf{\textit{Detect}} \\
        $\mathit{block_{ptr} = 0}$\\
        $\mathit{detect_{active} = 1}$
        }};

\path[->,every loop/.style={looseness=8}] 
	(I) edge [loop left, min distance=10mm] node { \textit{else}} (I)
	(D) edge [loop right, min distance=10mm] node { $\mathit{transfer_{last}}$} (D)
(M) edge [loop above, min distance=10mm] node { $\mathit{transfer_{inter}}$} (M);

\draw[->] (I) edge[above, bend left] node{ \shortstack{$\mathit{transfer_{inter}}$}} (M);
\draw[->] (M) edge[above, bend left] node{ \shortstack{$\mathit{transfer_{last}}$}} (D);
\draw[->] (M) edge[below, bend left] node{ \shortstack{$\mathit{detect_{any}}$ \\ $\lor$  \\ $~\mathit{transfer_{mismatch}}$}} (I);
\draw (D.south) -| (13.9,-3.5) -| (5.5,-3.5) node [below, bend left]{ \textit{else}} -| (0,-3.5) -> (I.south);
\draw[->] (D) edge[below, bend left] node{ \shortstack{$\mathit{transfer_{inter}}$}} (M);
\draw (I.north) -| (0,3.75) -| (5.5,3.75) node [above, bend left] {\shortstack{$\mathit{transfer_{last}}$}}  -| (13.9,3.75) -> (D.north);
\end{tikzpicture}
} 
\vspace{-1em}
\caption{State Machine of the Block Detection module(s)}
\vspace{-1em}
\label{fig:state-machine}
\end{center}
\end{figure}

The sub-path detection logic is defined in the finite state machine (FSM) depicted in  Fig.~\ref{fig:state-machine} with three states: $Idle$, $Monitor$, and $Detect$. Each Block Detect module is implemented as one instance of this FSM and is initialized in the $Idle$ state. All FSM transitions occur based on comparing the current entry in \cflog to $(\mathit{block_{src}, block_{dest}})$. Each Block Detect module maintains a pointer, $\mathit{block_{ptr}}$, which points to one ($\mathit{block_{src}, block_{dest}}$) pair in \data. For example, for a Block Detect module to monitor $(\mathit{block_{src_{i}}, block_{dest_{i}}})$, it sets $\mathit{block_{ptr}} = i$. Given the current $block_{ptr}$, a module can compare the current \cflog entry to the appropriate sub-path transfer to determine if the current \cflog entry:

\begin{compactitem}
\item  matches the first or an intermediate $(\mathit{block_{src}, block_{dest}})$ in the sub-path (denoted $\mathit{transfer_{inter}}$ in  Fig.~\ref{fig:state-machine});
\item matches the last $(\mathit{block_{src}, block_{dest}})$ in the sub-path (denoted $\mathit{transfer_{last}}$ in  Fig.~\ref{fig:state-machine}); or
\item is a mismatch (denoted $\mathit{transfer_{mismatch}}$ in  Fig.~\ref{fig:state-machine}).
\end{compactitem}
In addition, each Block Detect module receives the signal $\mathit{detect_{any}}$ from the outer \acron module, which is set when another Block Detect module has detected its sub-path.

When the FSM is in the $Idle$ state, $\mathit{block_{ptr}} = 0$ and ($\mathit{block_{src},}$ $\mathit{block_{dest}}$) contains the first transfer of the sub-path. The FSM transitions when the first ($src, dest$) occurs. If ($src, dest$) matches ($\mathit{block_{src},}$ $\mathit{block_{dest}}$) and the \vrf-specified sub-path contains just one control flow transfer, a transition to $Detect$ will occur because the whole path (one transfer) has been detected. Otherwise, a transition to $Monitor$ occurs. If ($src, dest$) does not match ($\mathit{block_{src}, block_{dest}}$), the FSM stays in $Idle$.


Upon entering $Monitor$, \acron has identified that the current \cflog entry matches the first transfer of the sub-path. To confirm a sub-path has occurred in \cflog, the Block Detect module must identify the exact remaining sequence of transfers in the sub-path. Therefore, while in $Monitor$, the Block Detect module increments $\mathit{block_{ptr}}$ for each match, to compare the next \cflog entry to the next transfer in the sub-path. For as long as $(\mathit{block_{src_i}}, \mathit{block_{dest_i}})$ matches \cflog entries, the FSM stays in the $Monitor$ state. If a mismatch occurs or another sub-path is detected ($\mathit{detect_{any}} = 1$), the FSM transitions back to $Idle$ and $\mathit{block_{ptr}}$ is reset to $0$. If the last transfer in the sub-path is reached, indicating a complete match, the FSM transitions to $Detect$.

In $Detect$, \acron hardware optimizes \cflog by replacing the sub-path with the $block_{ID}$. $\mathit{detect_{active}}$ is set to 1, triggering the Memory Interface module, which optimizes \cflog according to the $\mathit{spec_{addr}}$ and $\mathit{spec_{value}}$ values for the detected sub-path. After the replacement, $\ptr$ is updated to reflect the new reduced \cflog size. In addition, the $\mathit{block_{ptr}}$ is reset to 0. If the next transition is the first transfer of the sub-path, the FSM transitions to $Monitor$ or $Detect$ depending on the size of the sub-path. Otherwise, the FSM transitions back to the $Idle$ state.

\subsection{Hardware Specification Details}
\label{subsec:hw_specs}

\textbf{Memory Monitor:}
Per Fig.~\ref{fig:boundary}, this module monitors MCU signals to identify attempts to tamper with \data during the attested execution. 
To detect when the MCU is not executing the \CFA RoT, the Memory Monitor compares $PC$ to the bounds of the memory region storing the \CFA RoT code, denoted Trusted Computing Base (TCB).
If $W_{en}$ is set and $D_{addr}$ is within the bounds of \data, a CPU write to \data is occurring. 
If $PC$ is outside TCB, \acron interprets this action as an attempt by \adv to tamper with \data; thus, it triggers a hardware exception.
Similarly, if $DMA_{en}$ is set and $DMA_{addr}$ is within the bounds of \data, DMA is modifying \data (note that \data is a fixed reserved memory region). This triggers an exception at any time during execution.
Similar to other hardware exceptions on the target MCU (TI MSP430), this exception causes a system-wide reset. Note that similar controls from the underlying \CFA architecture~\cite{acfa} prevent unauthorized modification of \cflog.

\begin{figure}[t]
\scriptsize
\fbox{
    \parbox{0.95\columnwidth}{
        \underline{\textbf{HW Specification:}} Monitor Boundaries of \data
        \begin{equation*}
        \begin{split}
            &(\neg(PC \in TCB) \land (\mathit{W_{en}} \land (\mathit{D_{addr}} \in \data))) \\ 
            &\lor (\mathit{DMA_{en}} \land (\mathit{DMA_{addr}} \in \data)) \rightarrow reset \\
        \end{split}
        \end{equation*}
        \vspace{-1.5em}
    }
}
\caption{Memory Monitor specification}
\vspace{-1em}
\label{fig:boundary}
\end{figure}

\begin{figure}[t]
\scriptsize
\fbox{
    \parbox{0.95\columnwidth}{
        \underline{\textbf{HW Specification:}} Base address of Block$_i$ in \data:
        \begin{equation*}
            \mathit{block_{base_i}} =
            \left\{
            \begin{array}{ll}
                0, & \text{if } \mathit{i = 0}\\
                block_{base_{(i-1)}}+ (2\times block_{len_{(i-1)}}) + 1 & \text{otherwise}
            \end{array}
            \right.
        \end{equation*}
        \underline{\textbf{HW Specification:}} Given $\mathit{block_{base_i}}$ and $\mathit{block_{ptr}}$, get block signals
        \begin{equation*}
        \begin{split}
        &\mathit{block_{ID} := BlockMem[block_{base_i}][15:8]}\\
        &\mathit{block_{len} := BlockMem[block_{base_i}][7:0]}\\
        &\mathit{block_{src} := BlockMem[block_{base_i}+block_{ptr}+1]} \\
        &\mathit{block_{dest} := BlockMem[block_{base_i}+block_{ptr}+2]}
        \end{split}        
        \end{equation*}
        \vspace{-1em}
    }
}
\caption{  Block$_{i}$ Detect Module signals from \data}
\vspace{-1em}
\label{fig:blockmem_spec}
\end{figure}

\begin{figure}[t]
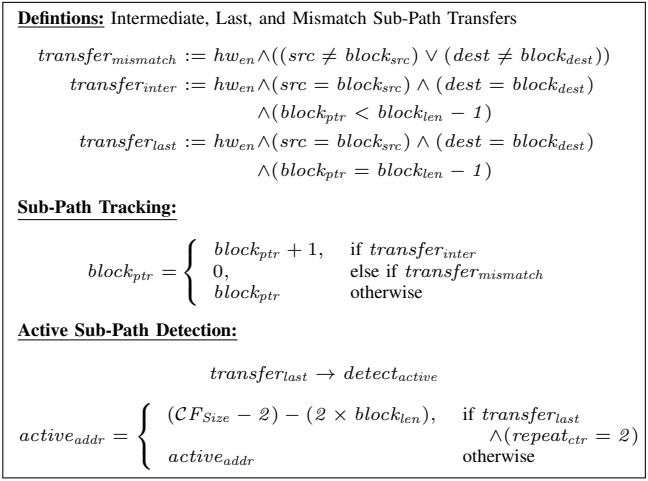

\scriptsize
\fbox{
    \parbox{0.95\columnwidth}{
    \underline{\textbf{Defintions:}} Intermediate, Last, and Mismatch Sub-Path Transfers
    \begin{equation*}
    \begin{split}
    \mathit{transfer_{mismatch} := hw_{en}} &\mathit{\land ((src \neq block_{src}) \lor (dest \neq block_{dest})}) \\
    \mathit{transfer_{inter} := hw_{en}} &\mathit{\land (src = block_{src}) \land (dest = block_{dest})} \\ &\mathit{\land (block_{ptr} < block_{len}-1)} \\
    \mathit{transfer_{last} :=  hw_{en}} &\mathit{\land (src = block_{src}) \land (dest = block_{dest})} \\ &\mathit{\land (block_{ptr} = block_{len}-1)}
    \end{split}\end{equation*}
    \underline{\textbf{Sub-Path Tracking:}}
    \begin{equation*}
        \mathit{block_{ptr} =}
        \left\{
        \begin{array}{ll}
            \mathit{block_{ptr}} + 1, & \text{if } \mathit{transfer_{inter}}\\
            0, & \text{else if } \mathit{transfer_{mismatch}} \\
            \mathit{block_{ptr}} & \text{otherwise}
        \end{array}
        \right.
    \end{equation*}
    \underline{\textbf{Active Sub-Path Detection:}}
    \begin{equation*}
    \begin{split}
        &\mathit{transfer_{last} \rightarrow detect_{active}}
    \end{split}
    \end{equation*}
    \begin{equation*}
        \mathit{active_{addr}} =
        \left\{
        \begin{array}{ll}
            \mathit{(\ptr - 2) - (2 \times block_{len})}, & \text{if } \mathit{transfer_{last}} \\ & \mathit{~~~~\land (repeat_{ctr} = 2)} \\
            \mathit{active_{addr}} & \text{otherwise}
        \end{array}
        \right.
    \end{equation*}
    \vspace{-1em}
    }
}
\caption{Block Detection specification}
\vspace{-1em}
\label{fig:ltl_bd}
\end{figure}

\textbf{Block Detect Module(s):} Each instance of this module abides by the state machine in  Fig.~\ref{fig:state-machine}.
The hardware specifications for interfacing with \data are shown in  Fig.~\ref{fig:blockmem_spec}.
Each Block Detect module must determine the sub-path block's base address ($\mathit{block_{base}}$) relative to the entire \data region to read the correct signals from \data.
This address is zero for the first sub-path block. The locations of all subsequent sub-path blocks depend on the length of the previous block(s), which are variable depending on \vrf's sub-path specification. Therefore, for the remaining blocks, the $\mathit{block_{base}}$ is calculated by incrementing the previous block's base address by its size. Each block contains a $\mathit{block_{ID}}$, $\mathit{block_{len}}$, and a series of control flow transfers. Since $\mathit{block_{len}}$ refers to the number of 32-bit control flow transfers in the sub-path, the size of the transfers in memory is $2\times block_{len}$ given 16-bit addresses in the target MCU. Both  $\mathit{block_{ID}}$ and $\mathit{block_{len}}$ are stored within a single address. As such $\mathit{block_{base_i}} = \mathit{block_{base_{(i-1)}}} + (2\times block_{len_{(i-1)}}) + 1$.

With $\mathit{block_{base_i}}$ and $\mathit{block_{ptr}}$, the remaining entries can be read from \data.
The first signal returned is $\mathit{block_{ID}}$, which references an 8-bit sub-path $\mathit{ID}$. The second signal is \textit{$block_{len}$}, which references the 8-bit length of the sub-path. Since MSP430 has 16-bit addresses, the upper bits of one address store \textit{$ID$}, while the lower bits store \textit{$len$}. Next, \textit{$block_{src}$} and \textit{$block_{dest}$} signals are set based on both the \textit{$block_{base_i}$} and \textit{$block_{ptr}$} values. Since the first address stored in a block is (\textit{$ID$} $|$ \textit{$len$}) (as shown in  Fig.~\ref{fig:component-details}), the first \textit{$(src, dest)$} are stored starting in an offset by 1 and 2 addresses, respectively. Therefore, to retrieve the expected \textit{$block_{src}$}, this module reads from \data at the location defined by \textit{$block_{base_i}+block_{ptr}+1$}. Similarly, the location defined by \textit{$block_{base_i}+block_{ptr}+2$} is referenced to return the expected \textit{$block_{dest}$}.

 Fig.~\ref{fig:ltl_bd} shows the hardware specifications for different types of transfers in a sub-path. With the signals from \data, each Block Detect module compares the current ($src, dest$) pair to (\textit{$block_{src},$} \textit{$block_{dest}$}) 
If the current \textit{($src, dest)$} does not match the current sub-path pair (\textit{$block_{src}, block_{dest}$}), then the current control flow transfer is identified as a mismatch ($\mathit{transfer_{mismatch}}$ bit is set). If the two pairs match, the module must determine if this is the last transfer in the sub-path or an intermediate transfer. The type of match is determined by comparing \textit{$block_{ptr}$} and \textit{$block_{len}$}. $\mathit{transfer_{inter}}$ is set when an intermediate sub-path transfer is matched. $\mathit{transfer_{last}}$ is set when the last sub-path transfer match is reached.
In accordance with the aforementioned logic, \textit{$block_{ptr}$} is incremented whenever an intermediate sub-path transfer match occurs and reset whenever a mismatch is reached.  

The internal signal $\mathit{transfer_{last}}$ will only be set when the last sub-path transfer has occurred, implying that all prior transfers were matches. At that stage, the output signal \textit{$detect_{active}$} is set.
The address of the active sub-path (\textit{$active_{addr}$}) is also updated with the start location of the sub-path in \cflog.
This address is determined by subtracting \textit{$2 \times block_{len}$} from the location of the last \textit{$(src, dest)$} pair in \cflog ($\ptr - 2$). 

\textbf{Memory Interface:} This module overwrites \cflog based on $spec_{value}$ and $spec_{addr}$. As described in Sec. \ref{subsec:comp_details}, $spec_{value}$ contains the ID of the detected sub-path, and $spec_{addr}$ holds the memory address in \cflog where the sub-path starts. When the $spec_{en}$ signal is set, $spec_{value}$ is written to \cflog at the memory address offset specified by $spec_{addr}$. Performing this write to \cflog makes $spec_{value}$ the last entry in \cflog. Therefore,
\ptr is updated to $spec_{addr} + 2$. Finally, in the specific case of multiple consecutive matches of the same sub-path, the Repeat Detect module is responsible for replacing repeated ID entries for a count of the number of repetitions
(see Appendix~\ref{rependix} for details).

\section{TEE-based \acron}
\label{sec:tz-design}

ARMv8 Cortex-M MCUs are equipped with TrustZone (TrustZone-M)~\cite{ARM-TrustZone}: an architectural security extension to create an isolated execution environment. Through this extension, hardware, software, and data on the MCU are divided into two worlds: ``Secure'' and ``Non-Secure'' Worlds. 
TrustZone ensures that Secure World code and data cannot be tampered with by code residing in the Non-Secure World (e.g., vulnerable/compromised MCU application code). Thus, security-critical data and code can be safely stored in the Secure World. 
Furthermore, the Non-Secure World can only call Secure World functions from secure entry points specified as Non-Secure-Callables (NSC), enabling controlled invocation of Secure World code from the Non-Secure World. 

\CFA techniques that leverage TrustZone-M~\cite{iscflat,oat,cflat} or other TEE support~\cite{scarr, recfa} leverage the Secure World to implement the \CFA RoT while untrusted application resides in the Non-Secure World. To record control flow transfers, the application is instrumented before deployment, and a call to the Secure World (via an NSC) is placed before each branching instruction. This NSC invokes a Secure World function for logging the branch destination to a Secure World \cflog. This design prevents unauthorized writes to \cflog.

\begin{figure}[t]
    \centering
    \includegraphics[width=\columnwidth]{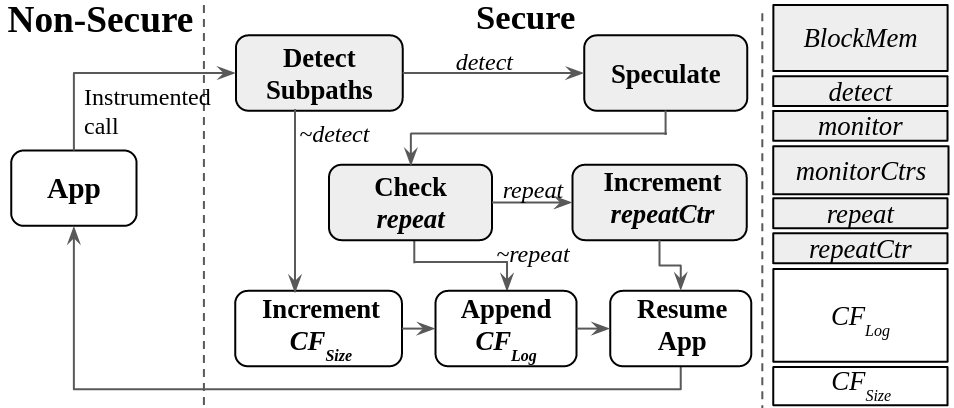}
    \vspace{-1.5em}
    \caption{\acron workflow on TrustZone-M}
    \vspace{-0.5em}
    \label{fig:tz-overview}
\end{figure}

Fig.~\ref{fig:tz-overview} presents the workflow of \acron in a TrustZone-based \CFA architecture. The \CFA RoT and \acron's functionality are implemented within the Secure World.
Each NSC switches to the Secure World to append \cflog with the new transfer. TrustZone makes the instrumented code unmodifiable between the point when it is measured (i.e., MAC-ed/signed by the \CFA RoT) and the point when the attested execution completes~\cite{iscflat}, guaranteeing to \vrf that all branches in the attested code are appropriately instrumented.

Without \acron, an NSC would invoke the \CFA RoT to append \cflog and resume the Non-Secure World execution.
\acron modifies this behavior to detect matching sub-paths. Similar to the custom hardware version (Sec.~\ref{sec:hw_based}), \data contains the definitions of \vrf's expected sub-paths, associated IDs, and sizes. For $n$ sub-paths, it maintains two $n$-bit values where each index corresponds to the \textit{detect} and \textit{monitor} signals of each monitored sub-path. Each bit signals when a particular sub-path is currently detected or monitored, respectively. For all sub-paths, a counter is stored in \textit{monitorCtrs}, holding the current number of consecutive transfers from that sub-path that have occurred in \cflog. For the most recently detected sub-path, a flag (\textit{repeat}) is maintained to signal when a sub-path repeats in adjacent \cflog locations, and a counter (\textit{repeatCtr}) stores the number of times it has repeated. All data is stored in the Secure World and thus is made inaccessible to a compromised application.

When an NSC triggers the \CFA RoT in \acron, it first uses the reported destination address to monitor and detect sub-paths. During this phase, the destination address is compared to the next address of each sub-path, determined from \textit{monitorCtrs} for each sub-path. When a match occurs for the next entry in any sub-path, the corresponding bit in \textit{monitor} is set, and its counter in \textit{monitorCtrs} is incremented. A sub-path match is detected if the new count equals the sub-path size, and the corresponding bit in \textit{detect} is set. Otherwise, \acron executes normally to append \cflog.

Once a sub-path is detected, \acron modifies \ptr to reduce \cflog and writes the sub-path \textit{ID} to \cflog. Then, it compares the current sub-path \textit{ID} to the last logged value. If they are equal, \acron increments \textit{repeatCtr} and returns to the application. When a sub-path is no longer repeating, its \textit{repeatCtr} is written to \cflog (conveying the number of consecutive repetitions of a sub-path).

\section{Security Analysis}\label{sec:security}

We argue that \acron added capability does not affect the security of the underlying \CFA architectures in producing a correct and authentic \cflog.
To see why, note that \adv (in the form of compromised software in \prv, as defined in Sec.~\ref{sec:adv_model}) may attempt to leverage \acron to launch attacks against \CFA in the following ways:
\begin{myenumerate}
\item \adv could attempt to modify \cflog directly.
If successful, this could cause \cflog to be reset and/or overwritten with entries that do not faithfully correspond to the executed control flow path.
\item \adv could attempt to corrupt the content of \data, modifying the \vrf-specified sub-path speculation to include the malicious transfers. A corrupted \data would cause \acron to optimize away the malicious path from \cflog, hiding the attack.
\item \adv could attempt to impersonate \vrf over the network to provide \prv with false sub-path speculations, hence overwriting \data to hide malicious sub-paths with symbols that would normally denote benign sub-paths.
\end{myenumerate}

Case 3 is prevented by ensuring that the \CFA RoT always authenticates \vrf requests (and sub-path speculations therein) before overwriting \data. Hence, the remainder of this analysis focuses on cases 1 and 2 for each type of architecture.

\textbf{\acron on Custom Hardware Architectures:}
In the hardware-based design of \acron, \cflog is protected against direct software writes by the underlying \CFA. Therefore, \adv must at run-time force an incorrect value of \textit{$spec_{addr}$}, \textit{$spec_{value}$}, or \textit{$spec_{en}$} to corrupt \cflog. Since \acron hardware controls these signals, they cannot be tampered with by any MCU software. Thus, they are unaffected by the untrusted software on \prv and \adv cannot corrupt \cflog. Unlike \acron signals, the contents of \data reside in the address space of the MCU. However, it is impossible for \adv to corrupt \data since the Memory Monitor in \acron prevents untrusted software modifications (by CPU or DMA) to \data.

\textbf{\acron on TEE-based Architectures:}
The TEE-based version of \acron stores \cflog in the Secure World. Therefore, \cflog is inaccessible to \adv in the Non-Secure World. Since the code of the attested program is unmodifiable during the attested execution and the application binary is instrumented (see Sec.~\ref{sec:tz-design}), modifications to \cflog can only occur through the instrumented NSC calls that follow a control flow transfer. Thus, \cflog is only appended with proper control flow transfer destinations, and \adv cannot reset/overwrite entries. Similarly, \adv cannot modify \data or \acron's implementation (code) because they are also stored in the Secure World. \data specifications (i.e., sub-path definitions) are received from \vrf as part of the \CFA request and authenticated by the \CFA RoT on \prv before the attested execution. Therefore, they are unmodifiable to the Non-Secure World.

\section{Selecting Sub-Path Speculations}\label{sec:selections}

Fundamentally, \acron performance depends on the effective selection of speculation sub-paths based on program and execution characteristics. As such, we examine multiple possible speculation strategies.

\subsection{Program Analysis}

Static analysis of the attested code can be used for sub-path speculation selection. This approach is valuable when \vrf lacks access to previous \cflog-s and must rely solely on program source code. Our static analysis-based approach examines both the \texttt{C} source code and the compiled program binary. Metadata from each function is collected, including the number of branching instructions, the number of loops, how many other functions call it, and how many times it calls other functions.

In our approach, we first extract the program's CFG. Once built, our implementation splits the CFG into "Segments," inspired by prior work in~\cite{zeitouni2017atrium}. Segments are determined by splitting the CFG into subgraphs at either the first node of the graph, the last node of the graph, the first node in a loop, or the last node in a loop. Splitting the CFG in this way ensures all Segments are forward-edge sub-graphs. Our implementation also splits the graph at calls and returns in order to avoid path explosion due to indirect calls and returns. Once the Segments are determined, the sub-paths within each Segment are enumerated and collected. After determining the Segment sub-paths for all functions, the set of Segments is optimized by combining those with just one successor Segment. The resulting set of candidate sub-paths is then sorted based on the following priorities:
\begin{myenumerate}
    \item sub-paths that exist within loops;
    \item sub-paths in the max-branching function;
    \item sub-paths in a function that is called within a loop or within the max-branching function in the code.
\end{myenumerate}
Sub-paths in functions that are never called or do not have any internal branches are not considered for the initial set of candidate sub-paths.

Based on the automatically generated candidate sub-paths, smaller sub-paths are first selected to optimize the utilization of \data in the initial set. After that, since the exact path that will occur might be highly unpredictable, non-overlapping paths are next selected to increase the initial coverage of the program. The sub-path selection can subsequently refined based on received \cflog-s.

\subsection{Automated \cflog Analysis} 
We also implement automated sub-path selection strategies that utilize past \cflog-s. Three policies were created to examine \cflog-s and recommend the best sub-paths: ``Top'', ``Minimize'', and ``Select''.

\textbf{Top} selects the most occurring non-overlapping sub-paths in the prior \cflog-s. 
However, \textit{Top} ignores sub-path sizes and can incur large memory overhead due to the increased size of \data to store sub-path speculations on \prv.

\textbf{Minimize} attempts to maximize \cflog reductions while minimizing \data sizes. It prioritizes small sub-path sizes first, then sub-path frequency. \textit{Minimize} chooses the $N$ most occurring smallest paths in \cflog, then iterates through all remaining sub-paths. Each remaining path is compared to the least-occurring selected path and replaces it if the candidate path occurs $t\%$ more frequently than the selected path, where $t$ is a configurable threshold. 

\textbf{Select} picks the most frequent sub-path that fits within the remaining memory in \data. It operates on the insight that, despite ``Minimize" reducing \data usage, the fixed-size \data remains part of \acron's memory overhead. Therefore, optimizing memory usage may involve filling the allocated space rather than minimizing \data use.

\subsection{Manual Inspection}

Another possible approach is to manually analyze previously received \cflog-s and the program's binary.
Developers may create custom speculations based on their insights into the attested program's behavior, expected inputs, or specific memory constraints.
While effective, this approach demands substantial human effort, limiting its scalability.
Moreover, as more \cflog-s are received, finding optimal speculation paths may become challenging.

\begin{table}[t]
\caption{Characteristics of Evaluated Applications}
\label{tab:baselines}
\vspace{-0.5em}
\centering
\begin{tabular}{|c|c|c|}
\hline
\textbf{App} & \textbf{Binary Size in Bytes} & \textbf{\cflog data (Bytes)}\\
\hline 
Ultrasonic~\cite{ultsensor} & 366 & 4160\\
Syringe~\cite{opensyringe} & 518 & 54600\\
Temperature~\cite{tempsensor} & 564 & 2508\\
Geiger~\cite{geiger} & 772 & 1740\\
Mouse~\cite{mouse} & 1119 & 50116\\
GPS~\cite{gps} & 6474 & 19876\\
\hline
\end{tabular}
\vspace{-1em}
\end{table}

\section{Prototypes and Evaluation}\label{sec:performance}

For \acron's version based on custom hardware, we use Xilinx Vivado tool-set~\cite{vivado} to synthesize \acron atop the ACFA architecture~\cite{acfa}, which targets the openMSP430 core~\cite{openmsp430}. \acron functionality was tested using the Vivado simulator to ensure its correctness. We then synthesized and deployed \acron on the Basys3 prototyping board, which features an Artix-7 FPGA. We implement the TEE-based version of \acron using a NUCLEO-L552ZE-Q development board equipped with an STM32L552ZE MCU which supports ARM TrustZone-M and is based on the ARM Cortex-M33 (v8) operating at 110 MHz.
We integrate \acron with the ISC-FLAT~\cite{iscflat} open-source TEE-based \CFA architecture. 
Both implementations use UART-to-USB as a communication interface with a baud rate of 38400.
%
To evaluate \acron on real MCU software, we port the open-source applications to both MSP430 and ARM Cortex-M. Table~\ref{tab:baselines} shows the evaluated applications and their characteristics.

\begin{figure}[t]
    \centering
    \subfigure[Added NSC Time of \acron-TZ\label{fig:logging_time}]{\includegraphics[width=0.49\linewidth]{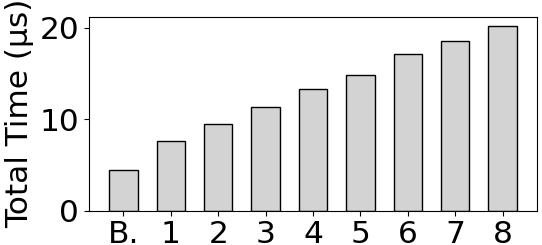}}
    \subfigure[Added cost of \acron-HW\label{fig:hardware}]{\includegraphics[width=0.49\linewidth]{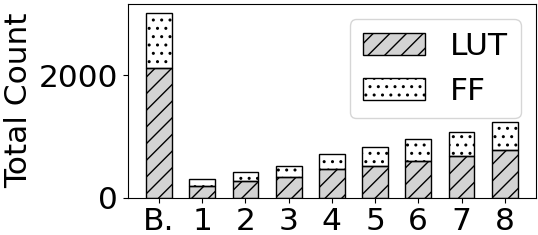}}
    \subfigure[\acron-HW vs. Related HW-CFA\label{fig:spec_vs_hw_cfa}]{\includegraphics[width=0.90\linewidth]{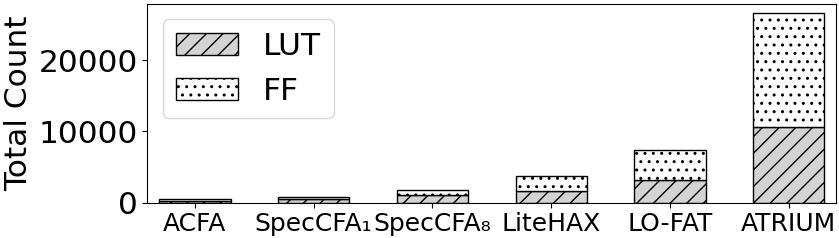}}
    \vspace{-0.75em}
    \caption{\acron cost analysis: (a) added NSC time (in $\mu s$) for 1-8 sub-paths atop baseline CFA logging; (b) additional HW cost of 1-8 sub-paths atop baseline of openMSP430+ACFA; (c) Comparison  to related HW-based \CFA.}
    \vspace{-0.75em}
\end{figure}
%
%
%

\subsection{\acron HW/Run-time Overheads}
The TEE-based version of \acron does not modify hardware (thus imposing no hardware overhead). However, it requires additional processing time for each \cflog entry, as matches are processed by the Secure World TCB (implemented in software). 
Thus, we assess the average time spent by NSC calls to the Secure World. Each NSC call must check the new entry against all active speculation sub-paths (until the first sub-path match -- or no match -- is found) to appropriately append to/optimize \cflog.
 Fig.~\ref{fig:logging_time} shows this cost as a function of the number of sub-paths checked for matches. On average, baseline \CFA without any speculation (B.) requires 4.4$\mu s$ to append an entry to \cflog. With 1 sub-path checked, the average time increases to 7.6$\mu s$. After that, each additional sub-path check adds 1.8$\mu s$, with 20.8$\mu s$ to check 8 independent sub-paths per NSC call. 

For the version of \acron based on custom hardware, similar to the related work~\cite{acfa,lofat,dessouky2018litehax,zeitouni2017atrium}, we measure the hardware overhead in terms of added Look-Up Tables (LUTs) and Flip-Flops (FFs). The increase in LUTs estimates the additional chip cost and size due to combinatorial logic, while the added FFs estimate the state overhead for sequential logic. We vary the number of supported speculation sub-paths from 1 to 8. The results are presented in  Fig.~\ref{fig:hardware}. As a conservative baseline, we show the cost of the openMSP430 core equipped with the underlying \CFA architecture~\cite{acfa}, excluding any hardware peripherals (e.g., general purpose I/O, communication interfaces, timers, etc.) that would normally add to the baseline hardware cost (B.) of the MCU. To support one sub-path speculation, \acron adds 190 LUTs and 107 FFs.
Each additional sub-path incurs an overhead of approximately 85 LUTs and 49 FFs.
The custom hardware-based design of \acron incurs no runtime overhead to speculate on sub-paths because its modules operate in parallel to the MCU core.

To put \acron's hardware overhead into context, we compare its cost with related \CFA architectures (that do not support configurable path speculations) in  Fig.~\ref{fig:spec_vs_hw_cfa}. As \acron implementation is built atop ACFA~\cite{acfa}, its relative cost increases accordingly. However, compared to other prior work in hardware-based \CFA (LiteHAX~\cite{dessouky2018litehax}, LO-FAT~\cite{lofat}, and ATRIUM~\cite{zeitouni2017atrium}), \acron incurs relatively low overhead. Even with support 8 sub-paths (``SpecCFA$_8$'' in  Fig.~\ref{fig:spec_vs_hw_cfa}), \acron's cost remains low in comparison. This indicates the feasibility of also deploying \acron on top of existing hardware-based \CFA at a relatively low overhead. Note that we could not implement \acron directly on top of these hardware-based \CFA architectures because they are not open-source. However, we see no reason why \acron design would not apply to them.


\subsection{Storage \& Communication Savings}
\label{subsec:memory_overhead}

\begin{figure}[t]
    \centering
    \subfigure{\label{fig:geiger_mem_overhead}\includegraphics[width=0.47\columnwidth]{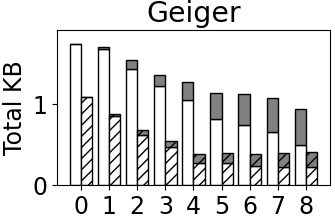}}
    \subfigure{\label{fig:ult_mem_overhead}\includegraphics[width=0.45\columnwidth]{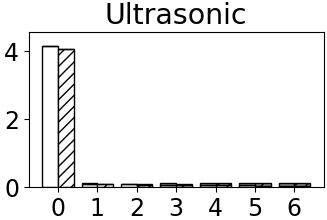}}
    \subfigure{\label{fig:temp_mem_overhead}\includegraphics[width=0.45\columnwidth]{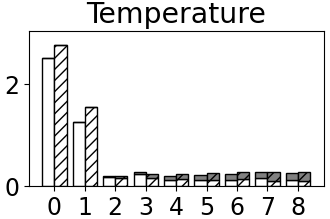}}
    \subfigure{\label{fig:syringe_mem_overhead}\includegraphics[width=0.45\columnwidth]{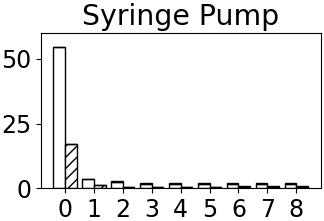}}
    \subfigure{\label{fig:gps_mem_overhead}\includegraphics[width=0.45\columnwidth]{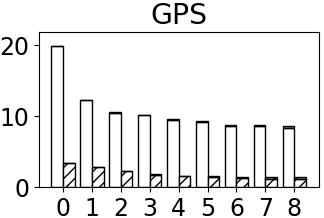}}
    \subfigure{\label{fig:mouse_mem_overhead}\includegraphics[width=0.45\columnwidth]{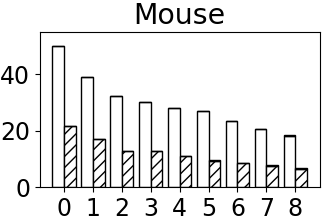}}
    \hfill
    \subfigure{\includegraphics[width=0.8\columnwidth]{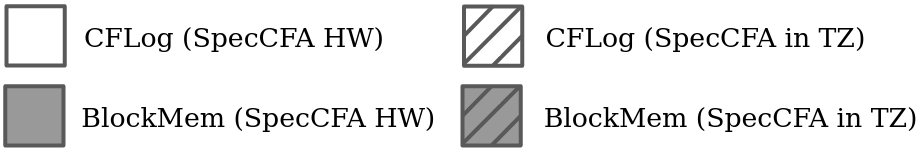}}
    \caption{\cflog and \data sizes (KB) for 1-8 simultaneous sub-path speculations.}
    \vspace{-1em}
    \label{fig:memory_overhead}
\end{figure}

To evaluate \acron's effectiveness, we start by selecting and configuring sub-paths based on manual inspection of the program source code and previously generated \cflog-s. In Sec.~\ref{sec:path_selection_results}, we revisit the automated methods discussed in Sec.~\ref{sec:selections}. We measure the overall reduction of \cflog for 1-8 sub-path speculations. We also contrast the reduced \cflog size with \data size (required to store the \vrf-defined speculation paths). This allows us to observe the total memory required to store both the optimized \cflog and respective sub-path specifications.
 Fig.~\ref{fig:memory_overhead} shows the memory requirement (in Bytes) to store \cflog and \data (stacked in each bar) for hardware-based and TEE-based \acron for 1-to-8 sub-paths.

In both designs, the total memory overhead is reduced as \acron speculates on more sub-paths.
Since the optimizations are application-aware, the savings vary across each application. For example, the first configured sub-path (selected as the most repetitive sequence of transfers) for Ultrasonic Sensor and Syringe Pump alone leads to significant savings. For these applications, optimizing based on 1 sub-path reduces the \cflog by 97.9-97.5\% and 93.3-92.2\%, respectively. This reduction is due to these applications executing many repeated control flow paths (e.g., repetitive iterations of signal processing functions and busy-wait loops). For the same reason, the Temperature Sensor gains most of its savings (94.4-93.1\% \cflog reduction) after speculating on two paths.
The GPS, Geiger Counter, and Mouse operations save at a steady rate as more sub-paths are configured. This increase in savings occurs because different sub-paths occur at similar rates.

Due to differences in the two underlying instruction sets (MSP430 vs. ARMv8 Cortex-M), the exact size of \cflog varies for each application in each architecture. \cflog-s for the TEE-based approach are on average smaller due to a more efficient instruction set and because instrumentation can allow the underlying CFA to ignore static branches. The hardware-based approach is implemented alongside MSP430, which has a reduced and less efficient instruction set. The hardware detects control flow transfers through the opcodes. Therefore, it records additional transfers for both direct and conditional jumps since they share the same MSP430 instruction opcode. We refer the reader to Appendix~\ref{apdx:impl_details} for additional architectural differences and implementation details. 

Naturally, the memory required to store \data increases as more sub-path speculations are used. However, the additional memory to store the sub-paths specifications is minimal compared to the savings made when speculating on them.

The memory savings shown in  Fig.~\ref{fig:memory_overhead} are also crucial for overall attested operation execution latency. Due to the limited amount of memory on \prv, execution may need to be interrupted to transmit a partial snapshot of \cflog to \vrf and free storage for additional transfers. This imposes significant delays to the attested execution, which can be avoided with \acron. We revisit this point in Sec.~\ref{sec:end-to-end_latency}.

\subsection{Path Selection Policy Comparison}\label{sec:path_selection_results}

\begin{figure}[t]
    \centering
     \subfigure{\label{fig:selections-geiger}\includegraphics[width=0.49\columnwidth]{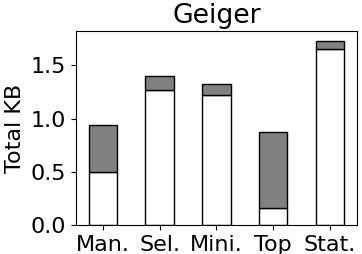}}
     \subfigure{\label{fig:selections-gps}\includegraphics[width=0.45\columnwidth]{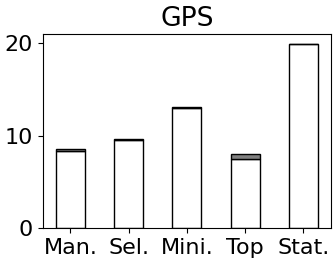}}
    \subfigure{\label{fig:selections-ultra}\includegraphics[width=0.45\columnwidth]{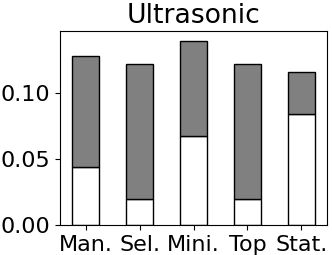}}
    \subfigure{\label{fig:selections-temp}\includegraphics[width=0.45\columnwidth]{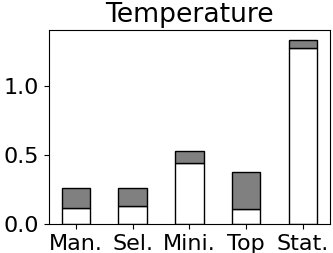}}
    \subfigure{\label{fig:selections-syringe}\includegraphics[width=0.45\columnwidth]{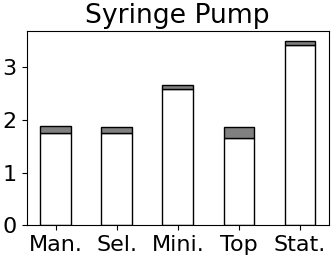}}
    \subfigure{\label{fig:selections-mouse}\includegraphics[width=0.45\columnwidth]{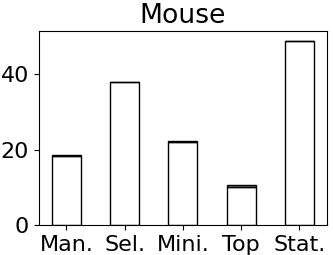}}
    \subfigure{\includegraphics[width=0.85\columnwidth]{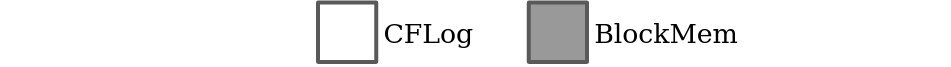}}
    \vspace{-.5em}
    \caption{\cflog and \data total size (KB) for each sub-path selection method: Manual Inspection (\textit{Man.}), Select (\textit{Sel.}), Minimize (\textit{Mini.}), \textit{Top}, and Static Analysis (\textit{Stat.)}}
    \vspace{-0.75em}
    \label{fig:selections}
\end{figure}

Fig.~\ref{fig:selections} shows the resulting memory overhead for each selection method as a sum of \data to store the suggested sub-paths and the optimized \cflog after speculation. Manual Inspection by a developer, in most cases, gains large \cflog optimizations due to prior knowledge of expected input/program behavior and recognizable patterns. However, this approach is not scalable and requires potentially impractical human efforts. Of the four automated processes, the static analyzer is the least performant. Since the static analysis does not consider any prior execution context, it is inherently limited in what it can learn about the program. However, it remains a suitable choice as a starting point when \cflog-s from prior executions do not yet exist or are unavailable. It performs best on programs that heavily execute repeated tasks, such as the Ultrasonic and Temperature Sensors.

All methods that inspect prior \cflog-s achieve optimizations comparable to or better than Manual Inspection. In most cases, \textit{Top} achieves comparable \cflog reductions to Manual Inspection. However, as it does not consider sub-path size, it normally incurs the largest \data overhead of the three \cflog inspection methods. \textit{Minimize} reduces the memory overhead associated with storing sub-paths. In prioritizing smaller sub-paths, \textit{Minimize} often misses better optimizations from longer sub-paths, decreasing \cflog reductions. Of the three methods, \textit{Select} achieves the most consistent optimizations, comparable to \textit{Top} while using less \data. For Geiger Counter, GPS, and Mouse, 
\textit{Top} beats \textit{Select} and \textit{Minimize} since the most occurring sub-paths in these programs have larger lengths. Similarly, \textit{Minimize} outperforms \textit{Select} for Mouse because \textit{Minimize} uses more \data than \textit{Select}.

\subsection{End-to-End Latency}\label{sec:end-to-end_latency}

To assess \acron's end-to-end effect on attested execution performance, we measure the total time taken to perform an attested execution of different operations on \prv.
In this case, a series of authenticated \cflog segments must be transmitted to \vrf throughout the execution of the application, whenever the \cflog's designated memory is full. This is required before subsequent transfers can be appended to \cflog. 
As a consequence, a significant portion of the attested execution time is spent on interruptions by the \CFA RoT to MAC/sign and transmit a \cflog slice to \vrf.

\begin{figure}[t]
    \centering    
    \subfigure[Custom hardware-based \acron]{
        \label{fig:msp-runtime}
        \includegraphics[width=0.85\columnwidth]{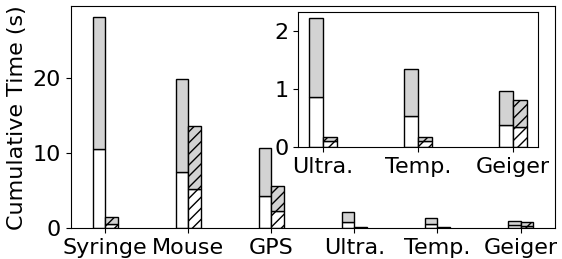}
    }
    \subfigure[TEE-based \acron]{
        \label{fig:tz-runtime}
        \includegraphics[width=0.8\columnwidth]{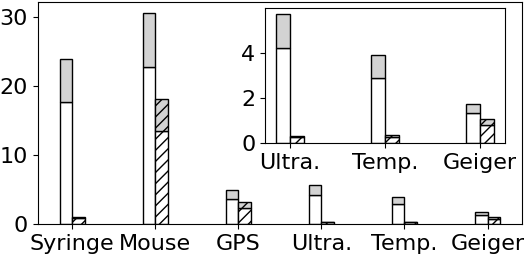}
    }
    \subfigure{\includegraphics[width=0.75\columnwidth]{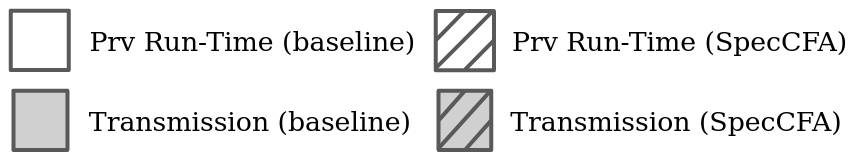}}
    \vspace{-0.5em}
    \caption{End-to-end latency of attested operations compared to baseline (without \acron).}
    \label{fig:runtime}
    \vspace{-1.5em}
\end{figure}

Given \acron's reduced \cflog sizes, fewer transmissions are required. To evaluate this impact on the overall \CFA performance, we measure the total attested execution time from when \vrf requests \CFA until when attested execution completes (including \vrf receipt of all execution evidence).  Fig.~\ref{fig:runtime} shows times of the custom hardware-based version and the TEE-based version and zooms in on Ultrasonic Sensor, Temperature Sensor, and Geiger Counter applications. 

In this experiment, \cflog slice sizes are set to 256 Bytes, and \acron is equipped to support 2 sub-paths. We then measure the total attested execution time (including the execution of the attested program, building \cflog, and generating a signature/MAC over \cflog). This total time is denoted ``\prv Run-Time''. We also measure the time spent transmitting the \cflog. Several applications present significant reductions in the overall attested execution time results. The Ultrasonic Sensor, Syringe Pump, and Temperature Sensor applications present $\approx$81-95$\%$ reduction in \prv Run-Time and $\approx$90-97$\%$ reduction in transmission. This performance improvement follows directly from the reduced amount of \cflog data that \prv must authenticate and transmit. With only 2 sub-paths, Geiger, Mouse, and GPS show less pronounced savings of $\approx$13-47$\%$ for run-time and $\approx$17-47$\%$ for transmission time. This follows from these applications requiring more than 2 sub-paths to exhibit pronounced \cflog reductions (recall  Fig.~\ref{fig:blockmem_spec}). 
Appendix~\ref{apdx:impl_details} contains additional implementation details of cryptographic operations and communication, as implemented by the underlying \CFA architectures~\cite{acfa,iscflat}.

\section{Limitations \& Potential Improvements}
\label{apdx:limitations}

The initial concept proposed in \acron presents several avenues for future work.

\textbf{Sub-path representation:} \acron's initial design represents and stores sub-path speculations verbatim on \prv. It would be interesting to propose other sub-path representations (e.g., using regular expressions or wildcards) that could enable 1 sub-path to be matched to multiple PMEM addresses based on offsets.

\textbf{Linear Hardware Cost:} In the initial design, the cost of \acron hardware increases linearly as one Block Detect module is required for each speculated sub-path. Therefore, future work could propose hardware optimizations to reduce \acron hardware overhead by serializing the sub-path detection and reducing the number of Block Detect modules.

\textbf{Additional Automated Sub-path Selection Methods}:
This work presented four methods for automated selection: one based on binary analysis and three based on inspecting prior \cflog-s. Future work should expand upon these methods to provide additional application-specific suggestions. Since our current program analysis is static, a potential future direction is to develop a system for dynamic analysis of a program and its sub-paths to gain insight into the likelihood of its execution. In addition, our current methods that inspect prior \cflog-s perform pattern matching. In future work, these methods can be extended to perform more advanced feature extraction from \cflog-s, prior input data, and the source code to make stronger predictions about possible future sub-paths.

\section{Conclusion}
We propose \acron: an approach to enable configurable application-aware sub-path speculations in \CFA. \acron provides \vrf with the ability to speculate on a program's likely sub-paths to reduce \cflog size significantly. Through \acron systematic design, \vrf can speculate on various sub-paths of any length without loss of information in \cflog. We implement two versions of \acron, based on custom hardware and on TEEs.
Our evaluation, performed on \acron's publicly available prototypes~\cite{specrepo}, demonstrates significant performance improvement for various MCU applications while retaining all standard \CFA security guarantees.

\section*{Acknowledgements}
We thank ACSAC'24 anonymous reviewers for their constructive comments. This work was partly funded by the National Science Foundation (SaTC award \#2245531).

{
\bibliographystyle{IEEEtran}
\bibliography{references}
}
    
\appendices


\section{Repeat Detect Module}
\label{rependix}

\begin{figure}[b]
  \centering
  \includegraphics[width=0.7\columnwidth]{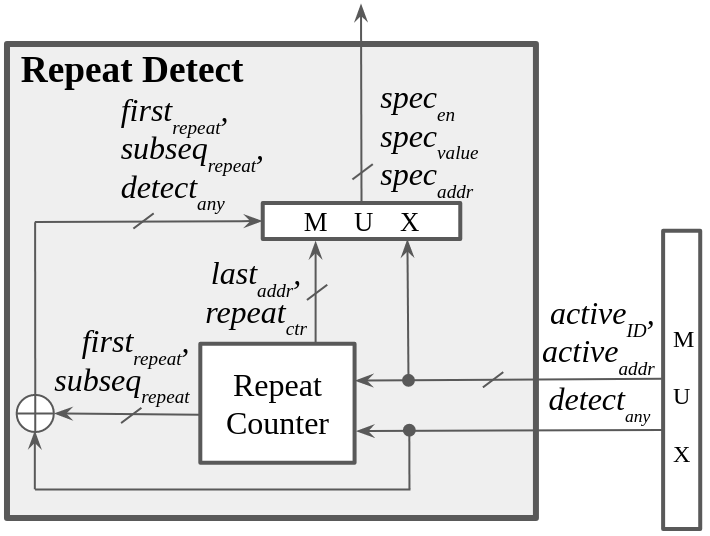}
  \caption{Repeat Detect module internals}
  \label{fig:rep_det_sys}
\end{figure}

 Fig. \ref{fig:rep_det_sys} depicts the Repeat Detect module in more detail. As discussed in Sec. \ref{subsec:comp_details}, the Repeat Detect module receives $detect_{any}$ indicating that a sub-path has been detected in \cflog. The module also receives \textit{$active_{ID}$} and \textit{$active_{addr}$} as an output from the MUX representing the ID of the occurring sub-path and its address in \cflog. Repeat Detect uses these signals to determine if a sub-path has repeated by comparing the current speculation to the previous one. For ease of representation, we depict this logic as the Repeat Counter in  Fig. \ref{fig:rep_det_sys}. If a repeat is detected, the address of the optimization \textit{$last_{addr}$} is determined, and a counter containing the number of consecutive repeats \textit{($repeat_{ctr}$)} is recorded. In addition, two signals are set ($first_{repeat}$ and $subseq_{repeat}$), whether this represents the first time the repeated sub-path occurred.

The final signals $active_{addr}$, $active_{ID}$, $last_{addr}$, and $repeat_{ctr}$ are passed to a MUX using the $\mathit{detect_{active}}$, $\mathit{first_{repeat}}$, and $\mathit{subseq_{repeat}}$ signals as a selector. The MUX determines the values of  $spec_{en}$, $\mathit{spec_{value}}$, and $\mathit{spec_{addr}}$ depending on if the detected sub-path is repeating. Upon the first occurrence of a sub-path, $\mathit{active_{ID}}$ and $active_{addr}$ are written to $\mathit{spec_{value}}$ and $spec_{addr}$. Then, if the sub-path repeats, they are set to $\mathit{repeat_{ctr}}$ and $last_{addr}$ instead. In both cases, $spec_{en}$ is set. Then, $spec_{en}$, $spec_{addr}$, and $spec_{value}$ are sent as output to the Memory Interface.

\subsection{Repeat Detect Hardware Specifications}

\begin{figure}[t]
\footnotesize
\fbox{
    \parbox{0.95\columnwidth}{
    \underline{\textbf{HW Specification:}} Select active block signals
    \begin{equation}
        \mathit{active_{ID} =}
        \left\{
        \begin{array}{ll}
            \mathit{block_{ID_0}}, & \text{if } \mathit{detect_{active_0}}\\
            \vdotswithin{3} &  \\
            \mathit{block_{ID_n}}, & \text{else if } \mathit{detect_{active_n}}\\
        \end{array}
        \right.
    \end{equation}
    \begin{equation}
        \mathit{active_{addr} =}
        \left\{
        \begin{array}{ll}
            \mathit{active_{addr_0}}, & \text{if } \mathit{detect_{active_0}}\\
            \vdotswithin{3} &  \\
            \mathit{active_{addr_n}}, & \text{else if } \mathit{detect_{active_n}}\\
        \end{array}
        \right.
    \end{equation}
    \begin{equation}
    \begin{split}
       &\mathit{(detect_{active_0} \lor ... \lor detect_{active_n}) \rightarrow detect_{any} }
    \end{split}        
    \end{equation}
    \underline{\textbf{Definitions:}}  Previous speculation address and ID
    \begin{equation}
        \mathit{last_{addr} =}
        \left\{
        \begin{array}{ll}
            active_{addr}, & \text{if } \mathit{detect_{any} \land (repeat_{ctr} = 2)}\\
            \mathit{last_{addr}} & \text{otherwise}
        \end{array}
        \right.
    \end{equation}
    
    \begin{equation}
        \mathit{last_{ID} =}
        \left\{
        \begin{array}{ll}
            active_{ID}, & \text{if } \mathit{detect_{any} \land (repeat_{ctr} = 2)}\\
            \mathit{last_{ID}} & \text{otherwise}
        \end{array}
        \right.
    \end{equation}
    \underline{\textbf{Definitions:}} Detect First and Subsequent Sub-Path Repeats
    \begin{equation}
    \begin{split}
    \mathit{first_{repeat} :=} &\mathit{(repeat_{ctr} = 2) \land (active_{ID} = last_{ID})} \\ 
    &\mathit{\land ((last_{addr}+2) = active_{addr})}
    \end{split}        
    \end{equation}
    
    \begin{equation}
    \begin{split}
    \mathit{subseq_{repeat} :=} &\mathit{(repeat_{ctr} > 2) \land (active_{ID} = last_{ID})} \\
    &\mathit{\land ((last_{addr}+2) = active_{addr})}
    \end{split}        
    \end{equation}
    
    \begin{equation}
    \begin{split}
    &\mathit{repeat := detect_{any} \land (first_{repeat} \lor subseq_{repeat})}
    \end{split}        
    \end{equation}
    \begin{equation}
        \mathit{repeat_{ctr} =}
        \left\{
        \begin{array}{ll}
            2, & \text{if } (\mathit{detect_{any} \land \neg repeat)}\\
            \mathit{repeat_{ctr}+1} & \text{else if } \mathit{detect_{any} \land repeat}
        \end{array}
        \right.
    \end{equation}
    
    }
}
 \caption{Hardware Spec: Repeat Detection}
\label{fig:repeat_detect}
\end{figure}

 Fig.~\ref{fig:repeat_detect} shows the additional hardware specifications for detecting when a sub-path repeats in adjacent \cflog locations. 
After a sub-path is detected and written to \cflog for the first time, \textit{$repeat_{ctr} = 2$}. At this moment, the internal value \textit{$last_{addr}$} is set in order to save the previous value of \textit{$active_{addr}$}, and \textit{$last_{ID}$} records the previous \textit{$active_{ID}$}. These values and \textit{$repeat_{ctr}$} are then used to detect when the first (\textit{$first_{repeat}$}) or subsequent repeats (\textit{$subseq_{repeat}$}) of the sub-path occur.

To detect the first repeat, the hardware checks if the next \textit{$active_{ID}$} is the same as \textit{$last_{ID}$} and if \textit{$repeat_{ctr}=2$}. If the IDs match, it then checks if \textit{$active_{addr}$} and \textit{$last_{addr}$} are adjacent by comparing \textit{$last_{addr}+2$} to \textit{$active_{addr}$}. If all three conditions are met, the first repeat has occurred. \textit{$repeat_{ctr}$} is then written to \cflog and incremented. To detect all subsequent repeats, a similar process occurs. If \textit{$last_{ID}$} and \textit{$active_{ID}$} are equal, \textit{$last_{addr}$} and \textit{$active_{addr}$} are adjacent, and  \textit{$repeat_{ctr} > 2$}, a subsequent repeat has occurred.

\begin{figure}[t]
\footnotesize
\fbox{
    \parbox{0.95\columnwidth}{
        %
        \underline{\textbf{HW Specification:}} Output speculation values
        \begin{equation}
            \mathit{spec_{en}} = detect_{any} \lor first_{repeat} \lor subseq_{repeat}
        \end{equation}
        \begin{equation}
            \mathit{spec_{value} =}
            \left\{
            \begin{array}{ll}
                \mathit{repeat_{ctr}}, & \text{if } \mathit{first_{repeat} 
           \lor subseq_{repeat}}\\
                \mathit{active_{ID}}, & \text{else if } \mathit{detect_{any}}\\
            \end{array}
            \right.
        \end{equation}
        \begin{equation}
            \mathit{spec_{addr} =}
            \left\{
            \begin{array}{ll}
                \mathit{last_{addr}+2}, & \text{if } \mathit{first_{repeat}}\\
                \mathit{last_{addr}}, & \text{else if } \mathit{subseq_{repeat}}\\
                \mathit{active_{addr}}, & \text{else if } \mathit{detect_{any}}\\
            \end{array}
            \right.
        \end{equation}
        %
        %
    }
}
\caption{Hardware Spec.: Interface with \data}
\label{fig:mem-interface}
\end{figure}

Repeat Detect outputs $spec_{en}$, $spec_{addr}$, and $spec_{value}$ to optimize \cflog. The hardware specifications for these signals are shown in  Fig. \ref{fig:mem-interface}. $spec_{en}$ is set as the logical OR of the signals \textit{$detect_{any}$}, \textit{$first_{repeat}$}, and \textit{$subseq_{repeat}$} as any of these signals indicate a sub-path has been detected.
When any repeat is occurring (\textit{$first_{repeat} \lor subseq_{repeat}$}), \textit{$repeat_{ctr}$} is used to set \textit{$spec_{value}$}. When no repeat occurs, and \textit{$detect_{any}$} is set, then \textit{$spec_{value}$} is set as \textit{$active_{ID}$}. Similarly, $spec_{addr}$ depends on these three signals (\textit{$detect_{any}, first_{repeat}, subseq_{repeat}$}). When the first repeat occurs, the counter must be logged to the address adjacent to the first sub-path ID, which is determined by \textit{$last_{addr}+2$}. On subsequent repeats, $last_{addr}$ contains the address of the counter in \cflog and can be overwritten directly with the new count. Finally, when there is no repeat but \textit{$detect_{any}$} is set, the address stored in \textit{$active_{addr}$} is used.
 
\section{Additional Implementation Details}
\label{apdx:impl_details}

\textbf{Platform differences effecting \cflog size:}
Compared to MSP430, ARMv8 Cortex-M has a richer instruction set. Because of this, some control flows can be optimized to make execution more efficient. For example, simple if statements and logical operations can be replaced with the execution of conditional instructions, removing certain control flow transfers.  We observe the effect of this difference in programs like Mouse and GPS. Several components of these programs, such as integer division and logical operations (such as setting a variable to the result of a comparison), are optimized in ARM, reducing the number of transfers. In addition, the STM32L552ZE MCU is equipped with an FPU, allowing floating point operations to require little to no control flow transfers. In MSP430, divisions require several branch instructions, and logical operations require control flow transfers. Hence, the \cflog reductions and selected sub-paths vary because some recurrent sub-paths in the control flow path of MSP430 binaries do not exist in ARMv8 binaries.

\textbf{Authenticating and Transmitting \cflog-s:}
Both implementations use UART-to-USB as a communication interface with a baud rate of 38400. The two use different cryptographic functions to authenticate \cflog-s. The underlying \CFA RoT~\cite{iscflat} for the TEE-based prototype uses a digital signature. It is implemented with SHA256 from HACL\text{*}~\cite{hacl} for hashing and Micro-ECC~\cite{micro_ecc_repo} for the Elliptic Curve Digital Signature Algorithm (ECDSA). It operates on an SECP256R1 curve with a $256$ bits private key to generate a $64$ Byte signature. For the custom hardware-based design, the underlying \CFA RoT\cite{acfa} uses SHA256-HMAC from HACL\text{*}~\cite{hacl} to produce a $32$ Byte MAC. Because of these differences, cryptographic operations consume most of the protocol time on the TEE-based version (due to the use of an asymmetric primitive) whereas transmission of data is the most time-consuming on the custom hardware-based version, as depicted in Fig~\ref{fig:runtime}.


\end{document}